\documentclass[letter]{aa} 

\usepackage{graphicx}
\usepackage{txfonts}
\begin{document}

   \title{Indications of a Si-rich bilateral jet of ejecta in the Vela SNR observed with {\em XMM-Newton}}

   \author{F. García\thanks{Fellow of CONICET, Argentina.}\fnmsep \inst{1,2} \and A. E. Suárez$^\star$\fnmsep\inst{1,2} \and M. Miceli\inst{3,4} \and F. Bocchino\inst{4} \and J. A. Combi\inst{1,2} \and S. Orlando\inst{4} \and M. Sasaki\inst{5}}

   \institute{Instituto Argentino de Radioastronom\'{\i}a (CCT-La Plata, CONICET; CICPBA), C.C. No. 5, 1894,Villa Elisa, Argentina
\and
Facultad de Ciencias Astron\'omicas y Geof\'{\i}sicas, Universidad Nacional de La Plata, Paseo del Bosque s/n, B1900FWA La Plata, Argentina. \email{fgarcia@iar-conicet.gov.ar, fgarcia@fcaglp.unlp.edu.ar}
         \and
         Dipartimento di Fisica e Chimica, Universit\`a degli Studi di Palermo, Piazza del Parlamento 1, I-90134 Palermo, Italy
         \and
             INAF-Osservatorio Astronomico di Palermo, Piazza del Parlamento 1, I-90134 Palermo, Italy
          \and
          Dr. Karl Remeis-Sternwarte, Erlangen Centre for Astroparticle Physics, Friedrich-Alexander-Universit\"at Erlangen-N\"urnberg,
Sternwartstrasse 7, D-96049 Bamberg, Germany
             }

   \date{Received ***; accepted ***}

 
  \abstract
   {The Vela supernova remnant displays several ejecta, which
are fragment-like features protruding beyond the front of its primary blast shock wave. They appear to be ``shrapnel'', bowshock-shaped relics of the supernova explosion. One of these pieces of shrapnel
(A), located in the northeastern edge of the remnant, is peculiar
because its X-ray spectrum exhibits a high Si abundance, in contrast to the other observed ejecta fragments, which show enhanced O, Ne, and Mg abundances.}
   {In this {\it Letter} we present the analysis of another ejecta fragment located opposite to shrapnel A with respect to the center of the shell, in the southwestern boundary of the remnant, named shrapnel G. We aim to fully characterize its X-ray emission to gather new information about the core-collapse supernova explosion mechanism.}
   {We thoroughly analyzed a dedicated {\it XMM-Newton} observation of shrapnel G by producing background-subtracted and exposure-corrected maps in different energy ranges, which we complemented with a spatially resolved spectral analysis of the X-ray emission.}
   {The fragment presents a bowshock-like shape with its anti-apex pointing to the center of the remnant. Its X-ray spectrum is best fit by a thermal plasma out of equilibrium of ionization with low O and Fe, roughly solar Ne and Mg, and a significantly high Si abundance, which is required to fit a very clear Si line at $\sim$1.85~keV. Its chemical composition and spectral properties are compatible with those of shrapnel A, which is located on the opposite side of the remnant.}
   {As a consequence of the nucleosynthesis, 
pieces of Si-rich shrapnel are expected to originate in deeper layers of the progenitor star compared to ejecta with lower-Z elements. A high velocity and density contrast with respect to the surrounding ejecta are necessary to make shrapnel A and G overtake the forward shock. The line connecting shrapnel A and G crosses almost exactly the expansion center of the remnant, strongly suggesting a Si-rich jet-counterjet structure, reminiscent of that observed in the young remnant Cas A.}


   \maketitle
%

\section{Introduction}

The explosion of a supernova triggered by the collapse of a massive star produces several solar masses of stellar ejecta expanding at $\sim$10$^4$~km~s$^{-1}$ into the surrounding circumstellar (CSM) and interstellar (ISM) material. The resulting forward shock compresses and heats the gas to high temperatures, thus  producing X-ray radiation. As the shock sweeps up material, the deceleration drives a reverse shock back into cold metal-enhanced ejecta, which are also heated to X-ray emitting temperatures. While in young historical supernova remnants (SNRs) the reverse shock is very close to the main blast wave and a significant fraction of the ejecta are still cold and unshocked, the reverse shock in evolved SNRs has had time to reach the SNR center, and therefore all the ejecta has been shocked and is expected to emit X-rays.

Several SNRs are characterized by a knotty ejecta structure, and very many clumps have been observed at different wavelengths in remnants of core-collapse supernovae (SNe), such as G292.0$+$1.8 \citep{park2004}, Puppis~A \citep{katsudaAPJ2008}, and Cas~A, where knots have also been detected beyond the main shock front \citep{hammelfesen2008,delaney2010}.

The Vela SNR represents a privileged target for studying the distribution of the ejecta and fragments detected beyond the forward shock front. It is considered to be the remnant of a Type~II-P SN explosion of a progenitor star with a mass lower than 25~M$_{\odot}$ \citep{gvaramadze1999}. Its age is estimated to be 11.4 kyr \citep{taylor1993}, and the distance to the SNR is about 250~pc \citep{bocchino1999,cha1999}. \cite{aschenbach1995} identified six ``shrapnel'' (labeled shrapnel A-F), which are X-ray emitting ejecta fragments with a characteristic boomerang shape protruding beyond the primary blast wave. Shrapnel A, B, and D have been studied in detail by \cite{tsunemi1999}, \cite{miyata2001}, \cite{katsudatsunemi2005,katsudatsunemi2006} and {yamaguchikatsuda2009}. These works have shown that the shrapnel may be divided into two categories. Shrapnel B and D have high O, Ne, and Mg abundances, while shrapnel A has a high Si abundance and weak emission from other elements. Consequently, \cite{tsunemikatsuda2006} have pointed out that the Si-rich shrapnel A must have been generated in a deeper layer of the progenitor than all the other shrapnel. Several bright ejecta knots have been discovered in the northern part of the remnant \citep{miceli2008}. The authors suggested that these knots would be shrapnel hidden inside the main shell by a projection effect, showing relative abundances similar to those found in shrapnel B and D.

In other core-collapse SNRs, the Si-rich ejecta may show a very peculiar jet-counterjet structure. Moreover, \cite{grichenersoker2017} suggested that jet-like features are common in many core-collapse SNRs. The well-known case of Cas~A has been studied in detail thanks to a very long {\it Chandra} observation \citep{hwang2004}
that shows a jet (with a weaker counterjet structure) composed mainly of Si-rich plasma. \cite{laming2006} have performed an X-ray spectral analysis of several knots in the jet and concluded that the origin of this interesting morphology lies in an explosive jet and it does not arise through interaction with a cavity or other peculiar structure of the ISM or CSM. Detailed 3D hydrodynamic simulations have shown that this jet can be explained as the result of velocity and density inhomogeneities in the ejecta profile of the exploding star with $\sim$2\% of the energy of the total energy budget of this remnant \citep{orlando2016}.

\begin{figure}
   \centering
   \includegraphics[trim={0 0 30 0},clip,width=0.47\textwidth]{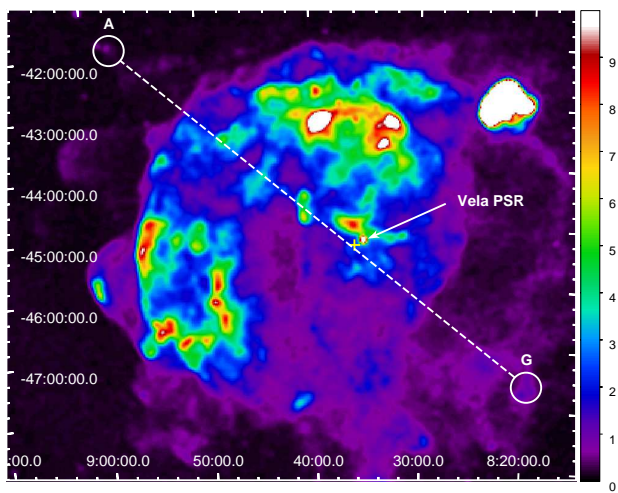}
      \caption{{\em ROSAT} All-Sky Survey image of the Vela SNR in the 0.44$-$2.04~keV energy range. Shrapnel A and G are indicated with white circles connected by a dashed line that crosses close to the Vela PSR, and the explosion point inferred from its age and proper motion (yellow cross).}
         \label{rosat}
   \end{figure}

In this {\it Letter} we present the analysis of an {\it XMM-Newton} dedicated observation of shrapnel G, located in the southwestern edge of the Vela SNR. We investigate the Si-rich ejecta in this shrapnel, which lies on the same line as the line that connects the center of the shell and the northeastern shrapnel A in the plane of the sky (see Fig.~\ref{rosat}). Shrapnel A and G reveal the signature of a jet-counterjet Si-rich structure in the ejecta of the very old Vela SNR, reminiscent of the very young core-collapse SNR Cas~A.

\section{Observations and data analysis}

The Vela shrapnel G has been observed once with the European Photon Imaging Camera (EPIC) of the {\it XMM-Newton} satellite. This camera consists of three detectors, two MOS cameras, namely MOS1 and MOS2 \citep{turner2001}, and a PN camera \citep{struder2001}, which operate in the 0.3$-$10 keV energy range. The observation was performed on April 22, 2012 (Obs. ID 0675080101), with a medium filter in Prime Full Window observation mode. The exposures were 42~ks, 43~ks, and 41~ks for the MOS1, MOS2, and PN cameras, respectively.

We analyzed the data using {\it XMM-Newton} Science Analysis System (SAS) version 15.0.0 and calibration files available in December 2016. In order to avoid soft-proton contamination, the observations were filtered using the SAS task {\sc espfilt} , resulting in reduced Good Time Intervals (GTI) of approximately 28ks for MOS1, 29ks for MOS2, and 27ks for PN exposures. For the subsequent analysis, the event lists were filtered to retain only events that likely stem from X-ray photons: we selected {\sc flag==0} events with single and double {\sc pattern} by means of the {\sc evselect} task.

\section{Results}

\subsection{X-ray images}

To produce images in different energy bands, we performed a double background subtraction to take into account particle and X-ray background contamination. For this purpose, we used Filter Wheel Closed and Blank Sky files available at {\it XMM ESAC} webpages\footnote{https://www.cosmos.esa.int/web/xmm-newton/filter-closed\\http://xmm-tools.cosmos.esa.int/external/xmm\_calibration/\\background/bs\_repository/blanksky\_all.html} and adopted the procedure described in \citet{miceli2017}. We then performed a point-source detection by running the {\sc edetect\_chain} script. Events in circular regions of 15~arcsec around each detected source were removed from the filtered event files, as we are only interested in the diffuse emission of the shrapnel.  We created background-subtracted images for each camera correcting for exposure and vignetting effects in different energy bands. Finally, we combined them by applying an adaptive smoothing by means of the {\sc emosaic} and {\sc asmooth} tasks. 

\begin{figure}
   \centering
   \includegraphics[trim={0 0 20 0},clip,width=0.47\textwidth]{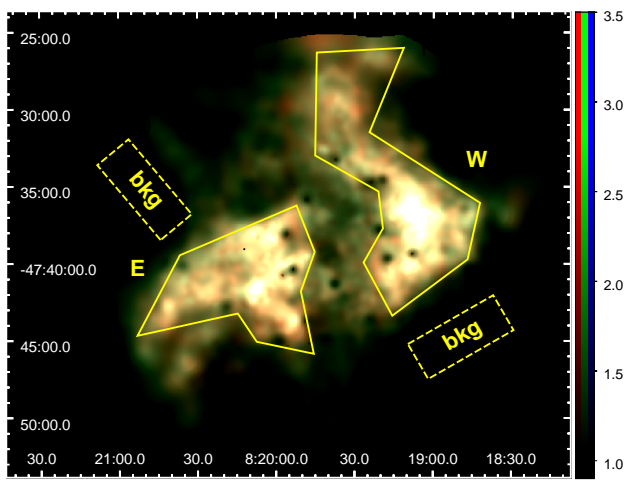}
      \caption{{\em XMM-Newton} false-color count-rate image of Vela shrapnel G. Red represents 0.3$-$0.6~keV, green 0.6$-$1.3~keV, and blue the 1.3$-$3.0~keV energy range. The overlaid yellow contours indicate the east (E) and west (W) spectral extraction regions. Dashed rectangles indicate the background regions (bkg). The image is background- and vignetting-corrected and point sources were removed.}
         \label{rgb}
   \end{figure}

In Fig.~\ref{rgb} we show the resulting composite X-ray image of the Vela shrapnel G obtained by combining the three EPIC exposures using a spatial binning of 4 arcsec. The soft band (0.3$-$0.6~keV) is shown in red, the medium band (0.6$-$1.3~keV) in green, and the hard (1.3$-$3.0~keV) band in blue. Black circles correspond to the subtracted point-sources. In the image, north is up and east is to the left. The shrapnel is indistinguishable from the background above 3.0~keV in the available data. As can be seen, the morphology of the shrapnel is fairly regular, showing two bright extended eastern (E) and western (W) regions. The X-ray emission shows a strong edge to the southwest (SW) coincident with a possible shock front of the shrapnel and weak elongated X-ray emission pointing to the northeast (NE), corresponding to the geometrical center of the Vela SNR.

\begin{figure}
   \centering
   \includegraphics[trim={0 0 20 0},clip,width=0.48\textwidth]{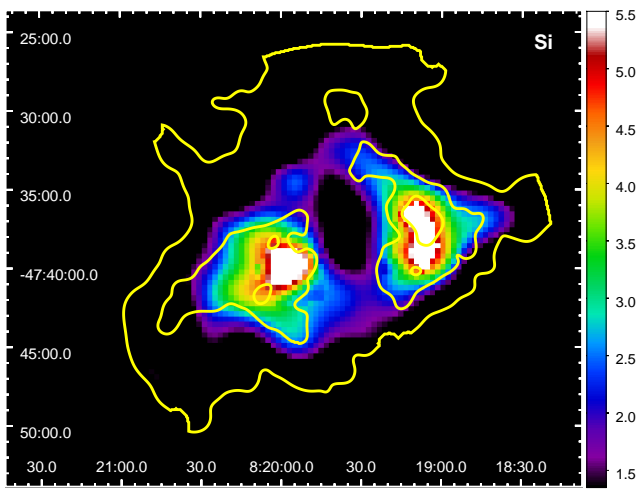}
      \caption{Background- and vignetting-corrected count-rate map in the 1.3$-$2.0~keV energy range, corresponding to the Si band. The overlaid yellow contours represent the emission in the 0.3$-$3.0~keV range. The Si maxima match the two bright knots of X-ray emission well.}
         \label{silicon}
   \end{figure}

In Fig.~\ref{silicon} we show a mosaiced X-ray map of the Si band (1.3$-$2.0~keV) with a spatial binning of 20 arcsec. The
overlaid yellow contours correspond to the X-ray emission in the whole 0.3$-$3.0~keV energy range. From this map it is evident that the photons originated in the Si band are spatially correlated with the total X-ray emission.

   \begin{figure}
   \centering
   \includegraphics[angle=-90,trim={0 30 0 0},clip,width=0.45\textwidth]{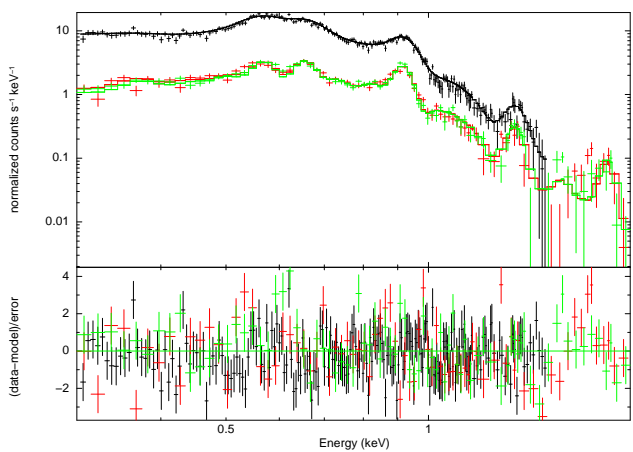}
      \caption{PN and MOS1/2 spectra of shrapnel G. Solid lines indicate the best-fit VNEI model (see Table~\ref{allspectable}). Lower panel shows the fit residuals.}
         \label{spectra}
   \end{figure}

\begin{table}
\caption{Spectral parameters of Vela shrapnel G.}
\renewcommand{\arraystretch}{0.9}
\begin{centering}
\begin{small}
\begin{tabular}{l | c | c}
\hline\hline
Model \& Parameters & {\bf VAPEC+VAPEC} & {\bf VNEI}  \\
\hline
$N_\mathrm{H}$ [10$^{22}$~cm$^{-2}$]    &       $0.11\pm0.01$ & $0.022\pm0.007$\\
\hline
$kT_{1}$ [keV]                          &       $0.194\pm0.002$         & $0.49\pm0.02$\\
Norm$_{1}$ [$\times 10^{-3}$]   &       $25\pm3$                & $2.0\pm0.3$\\
$\tau$ [$10^{10}$~s~cm$^{-3}$] &         $-$            & $3.1\pm0.3$\\
\hline
$kT_{2}$ [keV]                  &       $0.64\pm0.07$                   & $-$\\
Norm$_{2}$ [$\times 10^{-3}$]   &       $0.66\pm0.08$   & $-$\\
\hline
O,(=N),(=C)  [O$_\odot$]                &       $0.38\pm0.02$   & $0.47\pm0.05$\\
Ne [Ne$_\odot$]                         &       $1.14\pm0.08$   & $1.33\pm0.10$\\
Mg [Mg$_\odot$]                         &       $0.99\pm0.11$           & $0.92\pm0.12$\\
Si [Si$_\odot$]                         &       $2.06\pm0.45$           & $2.24\pm0.43$\\
Fe [Fe$_\odot$]                         &       $0.34\pm0.03$           & $0.29\pm0.04$\\
\hline
$\chi^{2}_{\nu}$ / d.o.f.       &       1.50 / 450                      & 1.41 / 451 \\ 
\hline
Flux (0.3$-$0.6~keV)            &       $4.49\pm0.06$ & $4.57\pm0.06$\\
Flux (0.6$-$1.3~keV)            &       $4.75\pm0.03$ & $4.73\pm0.03$\\
Flux (1.3$-$3.0~keV)            &       $0.33\pm0.02$ & $0.29\pm0.01$\\
\hline
Total Flux (0.3$-$3.0~keV)      &       $9.57\pm0.06$ & $9.59\pm0.06$\\
\hline
\end{tabular}
\end{small}
\label{allspectable}
\tablefoot{Normalizations are defined as 10$^{-14}/4\pi$D$^2\times {\rm EM}$, where ${\rm EM} = \int n_H\,n_e dV$ is the emission measure, $D$ is distance in [cm], $n_\mathrm{H}$ and $n_{\rm e}$ are the hydrogen and electron densities [cm$^{-3}$], and $V$ is the volume [cm$^{3}$]. Considering $D=250$~pc and a solid angle $\Omega=A/D^2=1.17\times10^{-5}$~sr for the spectral regions, we obtain an EM$/A=2.1\times10^{17}$~cm$^{-5}$ for the preferred VNEI model. Error values are 1$\sigma$ (68\%) confidence intervals for each free parameter. Fluxes are given in units of 10$^{-12}$~erg~cm$^{-2}$~s$^{-1}$ and solar abundances are taken from \cite{anders1989}.}
\end{centering}
\end{table}

\subsection{X-ray spectra}

Since the diffuse emission from the shrapnel G, which is an inhomogeneous extended source, occupies almost the full field of view, we applied the SAS {\sc evigweight} task to correct the event lists for vignetting effects. We obtained response and ancillary matrices using {\sc rmfgen} and {\sc arfgen} for a flat detector map, and we binned the spectra to obtain at least 25 counts per bin. We selected two polygonal spectral extraction regions named E and W, and background was extracted from two different regions where no significant emission from the shrapnel was detected (see Fig.~\ref{rgb}). The spectral analysis was performed using the XSPEC package \citep[Version 12.9.0,][]{arnaud1996} in the 0.3$-$2~keV band for the two MOS cameras and in the 0.3$-$1.5~keV band for the pn. The pn spectrum above 1.5~keV is dominated by the background, so we did not include it in our analysis to maximize the signal-to-noise ratio. We verified that our best-fit model does not change significantly by adding the pn data in the 1.5$-$2~keV band, although the indetermination of the best-fit parameters increases. Since we did not find a significant difference between the spectral fits of regions E and W, we combined them to improve the signal-to-noise ratio. Furthermore, we also checked that independently of the background region used, the best-fit values were consistent with each other. We inspected the background spectra and verified that they do not show any feature at the 1.85~keV Si band, which is visible only in the source spectra.

In Fig.~\ref{spectra} we show the background-subtracted X-ray spectra of the Vela shrapnel G for the three EPIC cameras (black for PN, red and green for MOS~1 and MOS~2, respectively). Errors are at 1$\sigma$ (68\%) confidence levels, and $\chi^{2}$ statistics are used. The X-ray spectrum of shrapnel G has a thermal origin, showing emission lines from \ion{O}{VII} (0.56~keV), \ion{O}{VIII} (0.65~keV), \ion{Ne}{IX} (0.92~keV), \ion{Mg}{XI} (1.35~keV), and Si (1.85~keV). We fit the spectra with different models of thermal emission from an optically thin plasma in collisional ionization equilibrium (APEC model) and in non-equilibrium of ionization (NEI model), taking into account the interstellar absorption \citep[PHABS,][]{balucinska1992}. The best fit is obtained for an NEI plasma with $kT=0.49\pm0.02$~keV. A two-temperature APEC plasma with $kT_1=0.194\pm0.02$~keV and $kT_2=0.64\pm0.07$~keV also gives a good fit. Non-solar abundances are required for both models to fit the emission lines, and their values are compatible between each other. The best-fit parameters are presented in Table~\ref{allspectable}. From now on, we focus on the single-temperature NEI model, which provides a significantly better fit to the spectra.

Before this study, a bright Si He$\alpha$ emission line could be detected only in the northeastern shrapnel A \citep{katsudatsunemi2006}. Remarkably, shrapnel G is located in the opposite southwestern region of shrapnel A. In contrast, shrapnel D \citep{katsudatsunemi2005} and the ejecta knots found by \cite{miceli2008} in the northern rim of Vela SNR showed no Si line and notably higher abundances of O, Ne, Mg and Fe than those found in shrapnel A \citep{katsudatsunemi2006} and G (this study). In particular, we found that in shrapnel G the Ne:Mg:Si:Fe:O abundances relative to O are 2.8:2.0:4.8:0.6:1 to be compared with 2.6:2.2:7.6:2.6:1 in shrapnel A, while shrapnel D and northern ejecta present Ne:Mg:Fe:O=2.1:2.2:0.2:1 and 2.5:3.2:0.5:1 and no Si line \citep{miceli2008}. Similar abundance patterns have also been observed by \cite{lamassa2008}, who found ejecta-rich plasma in the (projected) direction of the Vela PSR. In conclusion, shrapnel A and G are the only Si-rich shrapnel observed up to now in the Vela SNR. Furthermore, their plasma temperatures are consistent within 2$\sigma,$ and the ionization parameters are quite similar \citep[see Table~1 in][]{katsudatsunemi2006}, as are their projected sizes in the plane of the sky.

\section{Discussion}

In this {\it Letter} we presented a detailed study of the shrapnel G located in the SW region of the Vela SNR and showed the presence of Si-rich plasma in shrapnel G for the first time. We showed that shrapnel G has a very similar chemical composition to the geometrically opposite shrapnel A located in the NE edge of the remnant. While shrapnel B and D and all the other ejecta knots detected so far in Vela SNR have high O, Ne, and Mg abundances, shrapnel A and G have a high Si abundance and weak emission from other elements. 

As a consequence of the nucleosynthesis expected in stellar evolution, \cite{tsunemikatsuda2006} pointed out that Si-rich shrapnel must be generated in deeper layers of the progenitor star. However, hydrodynamic simulations of the Vela shrapnel show that an unrealistically high initial density contrast is required for an inner shrapnel to overcome outer ejecta knots, if we assume that the ejecta velocity increases linearly with their distance from the center \citep{miceli2008}. A possible solution for this issue is that (part of) the Si-burning layer has been ejected with a higher initial velocity, for example, as a collimated jet. This idea was later confirmed by dedicated 3D simulations of Cas~A, showing that both density and velocity inhomogeneities are necessary to reproduce the observed Si-rich jet \citep{orlando2016}.

The line connecting shrapnel A and G passes almost exactly through the expansion center of Vela SNR (see Fig. \ref{rosat}) as determined from the geometry of the other shrapnel and the proper motion of the Vela PSR. This alignment strongly supports the possibility that these shrapnel pieces are part of a Si-rich jet-counterjet structure. Assuming that the size along the line of sight, $L$, is equal to the projected size of shrapnel G in the plane of the sky (1320~arcsec) at a distance of $D=250$~pc, we estimate a number density of $n=0.21$~cm$^{-3}$ for the X-ray emitting plasma, and a total mass of $M=0.008$~M$\odot$ (for an average atomic mass of $2.1\times10^{-24}$~g for solar abundances). Considering the projected distance from shrapnel G to the geometrical center of the SNR and an age of $\sim$11~kyr, we obtain a velocity of $\sim$1400~km~s$^{-1}$ , leading to a total average kinetic energy of $E=1.6\times10^{47}$~erg, which is a lower limit when we take into account that the velocity of the shrapnel is not constant in time, that our estimate accounts only for the projected velocity, and that the X-ray emitting mass is only a fraction of the initial mass \citep{miceli2013}. Interestingly, the estimated mass of shrapnel G is similar (within a factor of 5) to the mass of the post-explosion anisotropy responsible for the Si-rich jet observed in Cas~A \citep{orlando2016}. On the other hand, our lower limit on the kinetic energy is two orders of magnitude lower than the
energy estimated soon after the SN explosion for the jet of Cas~A \citep[$\sim 4\times10^{49}$~erg,][]{orlando2016}. This is mainly due to our estimated velocity of shrapnel G: a velocity higher
by a factor of 10 (which is expected soon after the SN explosion) would produce energies similar to the energy found in the jet of Cas~A. We conclude that the mass and energy inferred for shrapnel G is very similar to the values estimated for the jet of Cas~A \citep{orlando2016}.

The structure of the ejecta in a SNR contains the imprint of the metal-rich layers inside the progenitor star. This type of detailed analysis of spatially resolved ejecta may help to understand the processes occurring in the latest stage of stellar evolution on the onset of core-collapse SNe explosions. In this sense, Vela SNR is an ideal candidate for performing this type of studies because of its age and angular size in the sky. Dedicated observations in the X-ray band of the remaining shrapnel that has not been studied so far are required to probe the still poorly understood physics of core-collapse supernovae and the formation of collimated ejecta jets.

With existing X-ray telescopes like {\it XMM-Newton}, it is possible to study small parts of a large SNR like Vela in pointed observations. Since we are not able to cover the entire SNR, it is difficult to achieve an understanding of the object as a whole. The German telescope {\it eROSITA} \citep[extended ROentgen Survey with an Imaging Telescope Array,][]{merloni2012} on board the Russian Spektrum-Roentgen-Gamma (SRG) mission, which is planned to be launched in 2018, will perform an all-sky survey (eRASS) in the 0.3$-$10~keV band for the first time. Equipped with CCDs similar to those of {\it XMM-Newton}, we will be able to study the entire SNR with a similar spatial and spectral resolution as we have presented here for shrapnel G. With a total exposure of $\sim$3~ks, eRASS will yield $\sim$23000 net counts for shrapnel G, allowing us to constrain abundances with an accuracy of $\sim$20\%.

\begin{acknowledgements}
We are grateful to the referee for very constructive comments. The research leading to these results has received funding from the European Union Horizon 2020 Programme under the AHEAD project (grant agreement n. 654215). FG and AES are grateful for the hospitality of INAF Osservatorio Astronomico di Palermo members. FG, AES and JAC were supported by PIP 0102 (CONICET). JAC was also supported by Consejer\'{\i}a de Econom\'{\i}a, Innovaci\'on, Ciencia y Empleo of Junta de Andaluc\'{\i}a under grant FQM-1343, and research group FQM-322, as well as FEDER funds. MM and SO aknowledge support by the PRIN INAF 2014 grant ``Filling the gap between supernova explosions and their remnants through magnetohydrodynamic modeling and high performance computing''. MS acknowledges support by the Deutsche Forschungsgemeinschaft (DFG) through the Heisenberg professor grant SA 2131/5-1.
\end{acknowledgements}

\end{document}